\documentclass{nature}

\usepackage{multibib}
\usepackage{graphicx}
\usepackage{amssymb,amsmath,bm,hyperref}

\usepackage{floatrow}
\DeclareFloatFont{tiny}{\scriptsize}
\def\aap{Astron.\ Astrophys.\ }
\def\apj{Astrophys.\ J.\ }
\def\apjl{Astrophys.\ J.\ Lett.\ }
\def\apjs{Astrophys.\ J.\ Supp.\ }
\def\aj{Astron.\ J.\ }
\def\mnras{Mon.\ Not.\ R.\ Astron.\ Soc.\ }

\def\prd{Phys.\ Rev.\ D\ }
\def\prl{Phys.\ Rev.\ Lett.\ }

\def\jcap{J.\ Cosmol.\ Astropart.\ Phys.\ }

\begin{document}

\begin{Large}
\noindent
\textbf{Insights from LHAASO and IceCube into the origin of the Galactic diffuse TeV--PeV emission}
\end{Large}

\noindent
{Kai Yan$^{1,2}$, Ruo-Yu Liu$^{1,2,*}$, Rui Zhang$^{3,4}$, Chao-Ming Li$^{1,2}$, Qiang Yuan$^{3,4}$, and Xiang-Yu Wang$^{1,2}$}

\begin{footnotesize}
\noindent
$^1$School of Astronomy and Space Science, Nanjing University, Nanjing 210023, China \\
$^2$Key laboratory of Modern Astronomy and Astrophysics (Nanjing University), Ministry of Education, Nanjing 210023, People's Republic of China\\
$^3$Key Laboratory of Dark Matter and Space Astronomy, Purple Mountain Observatory, Chinese Academy of Sciences, Nanjing 210023, China\\
$^4$School of Astronomy and Space Science, University of Science and Technology of China, Hefei 230026, China\\
*Email: ryliu@nju.edu.cn
\end{footnotesize}

\vspace{10pt}

\begin{abstract}
The high-energy diffuse gamma-ray emission and neutrino emission are expected from the Galactic plane, generated by hadronuclear interactions between cosmic rays (CR) and the interstellar medium (ISM). Therefore, measurements of these diffuse emission will provide important clues on the origin and nature of Galactic CRs. Comparing the latest observations of LHAASO and IceCube on the diffuse Galactic gamma-ray and neutrino emission respectively, we suggest that the diffuse gamma-ray emission at multi-TeV energies contains a considerable contribution of a leptonic component. By modelling the gamma-ray halos powered by middle-aged pulsars in our Galaxy with taking into account the magnetic field configuration and the interstellar radiation field in the Galaxy, we demonstrate that the collective contribution of pulsar halos can account for the excess in the measured diffuse gamma-ray emission with respect to the predicted flux from CR-ISM interactions.
\end{abstract}

The diffuse gamma-ray emission of the Galactic Plane (DGE) is the most prominent structure in the gamma-ray sky, as firstly revealed by satellites at the GeV band\cite{ Bignami75, EGRET97_DGE, Fermi12_DGE}. Ground-based TeV gamma-ray instruments can also measure part of the Galactic plane and the diffuse emission from the Galactic plane has been discovered at energies up to TeV-PeV band \cite{Milagro05, MILAGRO08, 2014HESSdiffuse, ARGO15_diffuse, Asgamma21_diffuse, 2023HAWCdiffuse} in recent decades. It is believed that the DGE mainly originates from hadronuclear interactions between Galactic cosmic rays (CRs) and the interstellar medium (ISM) \cite{Stecker73, Strong10, Lipari18}, with a possible contribution from unresolved, faint sources of either leptonic or hadronic emission. Therefore, the DGE contains critical information of the origin of Galactic CRs and extreme particle accelerators in our Galaxy. Very recently, the Large High-Altitude Air Shower Observatory (LHAASO) published the most complete TeV-PeV gamma-ray source catalog up to date in the declination between $-20^\circ$ and 80$^\circ$ \cite{LHAASO_catalog}, and the measurement of DGE in $10-1000$\,TeV after subtracting the contribution of these sources \cite{LHAASO_diffuse}. The measured flux, interestingly, is about 3 times higher than the prediction of CR-ISM interactions in the inner Galactic plane and 2 times higher than that predicted in the outer Galactic plane in the range of $\sim 10-60\,$TeV. The excess could be either caused by a CR overdensity with respect to the conventional model prediction due to additional complications in a realistic propagation model, or by contribution of unresolved sources.The latter scenario has been suggested to account for excesses in DGE measured by some other instruments \cite{Linden18, Liu21, ZhangPP22, Vecchiotti22, Yan23}, although the dominating source population has not been unambiguously identified. Either leptonic sources such as pulsar halos, or hadronic sources such as young stellar clusters may be responsible for the excess.

On the other hand, the IceCube neutrino telescope recently announced the detection of high-energy neutrinos originating from our Galaxy at a significance exceeding $4\sigma$ level with ten years of of cascade events data \cite{IceCube23}. The inferred Galactic neutrino flux accounts for approximately $8-12\,\%$ of the all-sky astrophysical flux at 30\,TeV. This discovery unambiguously serves as a smoking gun for the hadronuclear interactions between high-energy CRs and gas in the Milky Way. Although the poor angular resolution and limited statistics have prevented from identification of individual neutrino sources, the extracted neutrino flux, based on specific all-sky spatial templates, can still provide valuable insights into the hadronic component of the DGE. This article attempts to understand the origin of the DGE by leveraging the latest measurements from LHAASO and IceCube.

The neutrino analysis is based on theoretical all-sky spatial templates, while the DGE flux given by LHAASO is extracted from a portion of the Galactic plane ($|b|<5^\circ$ and $15^\circ<l<235^\circ$), after masking all known sources and their surrounding regions. Furthermore, the neutrino emission measured by IceCube also likely contains a considerable contribution from neutrino sources in the Galaxy \cite{IceCube23}, in addition to those truly diffusive emission from CR-ISM interactions. Consequently, direct comparison of the results from the two measurements is not feasible.

In the first step of the analysis, we aim to assess the proportion of the observed neutrino flux contributed by hadronic gamma-ray sources in the Galaxy, in order to obtain the diffuse neutrino flux related to CR-ISM interactions. As a ground-based gamma-ray detector, LHAASO cannot detect most sources in the southern hemisphere. We therefore combine sources reported in the first LHAASO catalog with $|b|<5^\circ$ and sources recorded in the Galactic Plane survey of High Energy Spectroscopic System (HESS) with declination less than $-20^\circ$, in order to compile a more complete source list covering the entire Galactic plane. To exclude gamma-ray sources of leptonic origin, we omit sources that are spatially associated with pulsars, despite suggestions that pulsars or pulsar wind nebulae may also be high-energy proton accelerators \cite{Atoyan96, Amato03, Liu21_crab}. On the other hand, we find that the neutrino flux from the Galactic plane within $|b|<5^\circ$ is about 50\% of the all-sky flux according to the Fermi-LAT $\pi^0$ template \cite{Fermi16_DGE_template}, and about 70\% of the all sky flux according to the KRA$_\gamma^5$ template \cite{Gaggero15_DGE}, which are employed in IceCube's data analysis. The gamma-ray flux and the single-flavor neutrino flux generated in the same hadronuclear interactions are approximately related by
$E_\gamma^2\frac{dN_\gamma}{dE_\gamma}=2E_\nu^2\frac{dN_\nu}{dE_\nu}$,
where $E_\gamma=2E_\nu$. We then can convert the neutrino flux in the Galactic plane to the corresponding gamma-ray flux and compare it with the total source flux. As shown in Figure~1, the sum of the source contribution, despite the systematic uncertainties in the source fluxes\footnote{Considering the systematic uncertainty would make the ratio between the source flux and the total flux vary in the range of $\sim 1/4-1$.}, can reach about 2/3 of the total flux converted from the neutrino measurement with the $\pi^0$ template, from a few TeV up to several tens of TeV. Although the measured neutrino flux is based only on the template for CR-ISM interactions, the contribution from sources would increase the normalization of the inferred neutrino spectrum significantly. Beyond $\sim 100$\,TeV, on the other hand, we can see the source contribution is less important, implying that Galactic neutrinos may be dominated by CR-ISM interactions above this energy. The neutrino flux in the Galactic plane obtained with the KRA$_\gamma^5$ template is even slightly lower than the source contribution, which is not physical. Actually, the KRA$_\gamma^5$ template gives rise to a lower significance of neutrino detection than the $\pi^0$ template, which may not reflect the neutrino flux level in the Galaxy accurately. Nevertheless, it also indicates a significant contribution of sources to the neutrino emission in the Galactic plane.  We note that a previous study \cite{Fangke23_na} proposed that the neutrino flux of the Galaxy measured by IceCube is consistent with that converted from the DGE flux measured by AS$\gamma$. However, the DGE flux measured by AS$\gamma$ is largely contaminated by emissions from Galactic sources because of underestimation of source extensions and an incomplete source list employed in their source-masking procedure \cite{Liu21}. This is evidenced by the difference between the flux measured by AS$\gamma$ and that by LHAASO as shown in Figure~2, where we can see the former is $4-5$ times higher than the latter around 100\,TeV. This supports our finding here that a large fraction of the measured neutrino flux of the Galaxy originates from individual Galactic sources.

\begin{figure*}[htbp]
\centering
\includegraphics[width=0.5\textwidth]{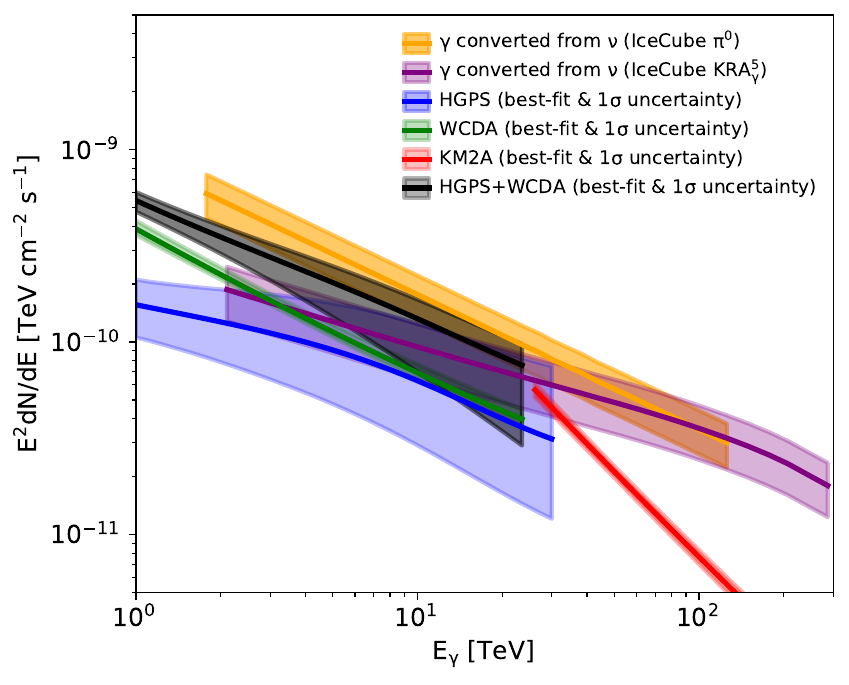}\\
\caption{Comparison of the gamma-ray flux derived from neutrino measurement and the total flux of known Galactic gamma-ray sources excluding those with pulsar association. The orange and purple curves show the gamma-ray fluxes in the Galactic plane ($|b|<5^\circ$) derived from the neutrino measurement with the Fermi-LAT $\pi^0$ template and KRA$_\gamma^5$ template, respectively. The green and red curves show the total gamma-ray fluxes of sources measured by the Water Chenrenkov Detector Array (WCDA) and Kilometer Square Arrary (KM2A) of LHAASO detectors, respectively. The blue curve shows the total flux of source with declination less than $-20^\circ$ measured in the HESS Galactic Plane Survey (HPGS). The black curve presents the sum of measurements from WCDA and HESS. Shaded regions of corresponding colors represent $1\sigma$ uncertainties of the fluxes, including both statistical errors and  systematic errors.}
\label{fig:multi_flux}
\end{figure*}

Next, to estimate the hadronic gamma-ray component in the DGE measured by LHAASO, we need to find out the neutrino flux from the same region of interest (ROI) as LHAASO's DGE analysis. LHAASO's DGE analysis masked each of the known sources and the surrounding ISM in the sky map to reduce the contamination of the source to the measured DGE. Therefore, we also need to mask the same region in the all-sky template for the neutrino analysis. We find that the flux ratio of the remaining region to the all-sky map is 12.5\% for the Fermi-LAT $\pi^0$ template and 8.1\% for the KRA$_\gamma^5$ template  (see Extended Data Figure~1). Dividing the rescaled neutrino flux by the corresponding solid angle of the remaining region, which is 0.206\,sr for the inner Galactic plane ($|b|<5^\circ, 15^\circ<l<125^\circ$) and 0.268\,sr for the outer Galactic plane ($|b|<5^\circ, 125^\circ<l<235^\circ$), we can obtain the average neutrino intensity, which is supposed to be related to CR-ISM interactions, from the same ROI of LHAAOS's DGE measurement. After converting the neutrino intensity into the corresponding gamma-ray intensity, we can compare it to LHAASO's DGE measurement.

We note that some uncertainties may exist in the above analysis. First, we may expect the existence of some unresolved hadronic TeV gamma-ray sources in the Galaxy. Considering contributions of these sources would further reduce the gamma-ray flux related to CR-ISM interactions.  This could be in particular important for sources in the southern hemisphere, because of the relatively low sensitivity of HESS Galactic Plane Survey (HGPS) above 10\,TeV (See Supplemental Material). On the other hand, some of known gamma-ray sources without pulsar association could be actually of the leptonic origin. Indeed, the hadronic nature of some gamma-ray sources, such as Cas A, RX J1713.7-3946, is still under debate. However, most of this kind of sources shows early cutoff/steepening below $\sim 1-10$\,TeV in their gamma-ray spectra \cite{Aharonian19}. Even if we regard emissions of these sources as leptonic, we do not expect a significant decrease of the total hadronic source flux above 10\,TeV, which is the energy range relevant with LHAASO's DGE measurement. The uncertainties from these two aspects operate in the opposite direction, and may cancel each other to a certain extent. An accurate evaluation of the uncertainty relies on further studies of these sources, via multiwavelength and multimessenger observations.

\begin{figure*}[htbp]
\centering
\includegraphics[width=\textwidth]{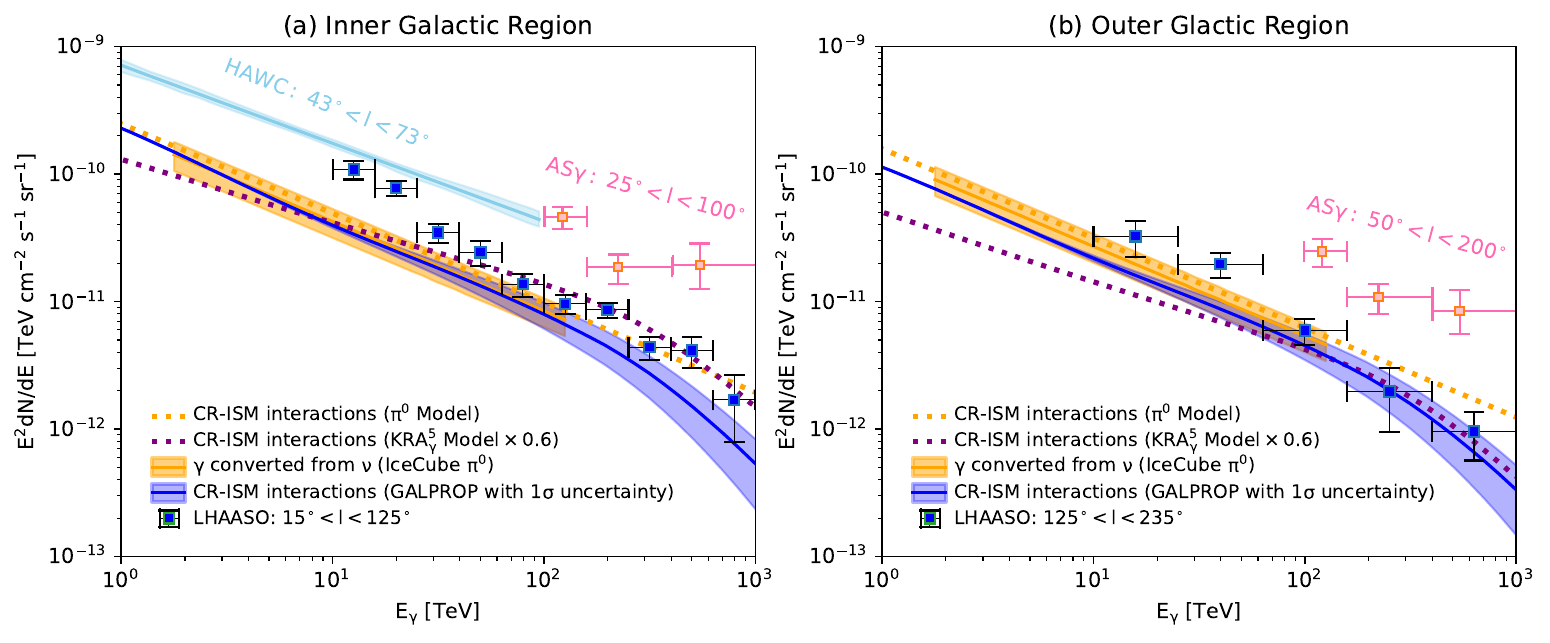}
\caption{Comparison of the DGE fluxes derived from different analyses. (a) the inner Galactic plane. (b) the outer Galactic plane. The orange shaded areas show the DGE of hadronic origin derived from neutrino measurement with the Fermi-LAT $\pi^0$ template. The blue shaded areas show the predicted flux of CR-ISM interactions (blue shaded region indicating the 1$\sigma$ uncertainties), which follows the calculation in Ref.~\cite{ZhangR23}. The dotted orange and purple curves show the predicted flux of CR-ISM from the Fermi-LAT $\pi^0$ model and the KRA$_\gamma^5$ model, respectively. The curve for KRA$_\gamma^5$ model is rescaled by a factor of 0.6 with respect to the original value in order to match the LHAASO data above 100\,TeV. The blue squares represent the DGE flux measured by LHAASO\cite{LHAASO_diffuse}. For reference, the DGE reported by AS$\gamma$\cite{Asgamma21_diffuse} and HAWC\cite{2023HAWCdiffuse} are also shown as pink squares and lightblue band, respectively.}
\label{fig:ic_flux}
\end{figure*}

Finally, we present the comparison between the gamma-ray flux derived from the neutrino measurement and the DGE measured by LHAASO in Figure~2. We see that the derived gamma-ray intensity related to CR-ISM interactions is lower than the LHAASO measured DGE flux below $\sim 100\,$TeV for both the inner Galactic region and the outer Galactic region. The predicted diffuse gamma-ray intensity from CR-ISM interactions (blue lines and shaded regions), following the calculation by Zhang et al. \cite{ZhangR23}, is shown for reference. This calculation employs GALPROP \cite{GALPROP11} and is based on up-to-date measurements of
the local CR spectra from various instruments for CR protons, helium nuclei and heavier isotopes such as Li, Be, B, C, O, as well as electrons and positrons \cite{ZhangR23}. We see that an excess in the DGE measured by LHAASO is obvious below $\sim 60$\,TeV with respect to the model prediction. For comparison, we also show the diffuse emission predicted by the $\pi^0$ model the $\rm KRA_{\gamma}^5$ model in the ROI of LHAASO's DGE analysis after performing the same mask procedure. The DGE flux predicted by the $\pi^0$ model (orange dotted lines in Figure~2) is consistent with the prediction by GALPROP. On the other hand, the flux predicted by the $\rm KRA_{\gamma}^5$ model slightly exceeds the data above 100\,TeV. By normalising its flux to the data at $100\,$TeV, an excess of the data with respect to the model prediction can be also seen below $\sim 50\,$TeV.
The excess indicates an additional component contributing to the DGE, and must be of the leptonic origin. A plausible candidate population for the leptonic DGE excess is pulsar halos, which have been discussed as potential contributors to DGE \cite{Linden18, 20Cataldo, Vecchiotti22, Martin22, Yan23, Dekker23} with simple modelings of the contribution of the pulsar halo population. In what follows, we will explore in detail the pulsar halo interpretation for the DGE excess measured by LHAASO.

Pulsar halos are produced by the IC scattering of energetic electron/positron pairs, which have escaped from their pulsar wind nebulae (PWNe), off the background radiation field in the ISM. From the perspective of energy budget, pulsar halos are powered by the rotational energy of pulsars, with a fraction $\eta_e$ of the pulsar's spindown power converted into energetic pairs and subsequently into radiations. Pulsar halos are apt to form around pulsars with age beyond several tens of thousand years \cite{Giacinti20}, as the confinements of pairs are relatively weak inside these middle-aged PWNe while the spindown powers of the pulsars are not very low. Although the transport mechanism of injected pairs in the surrounding ISM is still under debate, a consensus has been reached that pulsar halos are extended TeV gamma-ray sources without sharp boundaries. It is therefore natural to expect that many pulsar halos are faint, diffusive gamma-ray sources and most of them are not resolved by instruments.

To assess their contribution to the DGE, we single out pulsars with characteristic age between 50\,kyr and 10\,Myr from the ATNF pulsar catalog \cite{Manchester05}. Here we consider two propagation models for escaping pairs. The first one is the two-zone isotropic diffusion model (2ID model) with a suppressed diffusion coefficient $D_0(E)$ within a radius of $r_b$ from the central pulsar and the typical diffusion coefficient of the ISM $D_{\rm ISM}$ as inferred from the secondary-to-primary CR ratio for the region beyond $r_b$ \cite{Fang18, Profumo18}. In the second model, we consider the anisotropic diffusion model (AD model) \cite{Liu19_prl} in which particle diffuse more rapidly along the mean magnetic field direction than they diffuse perpendicular to the mean field direction. In both models, we generate the interstellar magnetic field strength $B_0$ in the vicinity of each pulsar based on their positions in the Galaxy and the Galactic magnetic field (GMF) model suggested by Jansson \& Farrar \cite{JF12_regular, JF12_random} (see Extended Data Figure~2), and consider the interstellar radiation field density following the model proposed by Popescu et al. \cite{Popescu17}. As such, we can calculate cooling and radiation of pairs around each pulsars. For the AD model, we also generate the mean direction of the magnetic field because it determines the preferential diffusion direction of injected pairs. The turbulent level of the magnetic field, indicated by the Alfv{\'e}nic Mach number $M_A=\Delta B/B_0$, is then randomly assigned in the range of $(0.1, 1)$ which is typical for ISM.

\begin{figure*}[htbp]
\includegraphics[width=\textwidth]{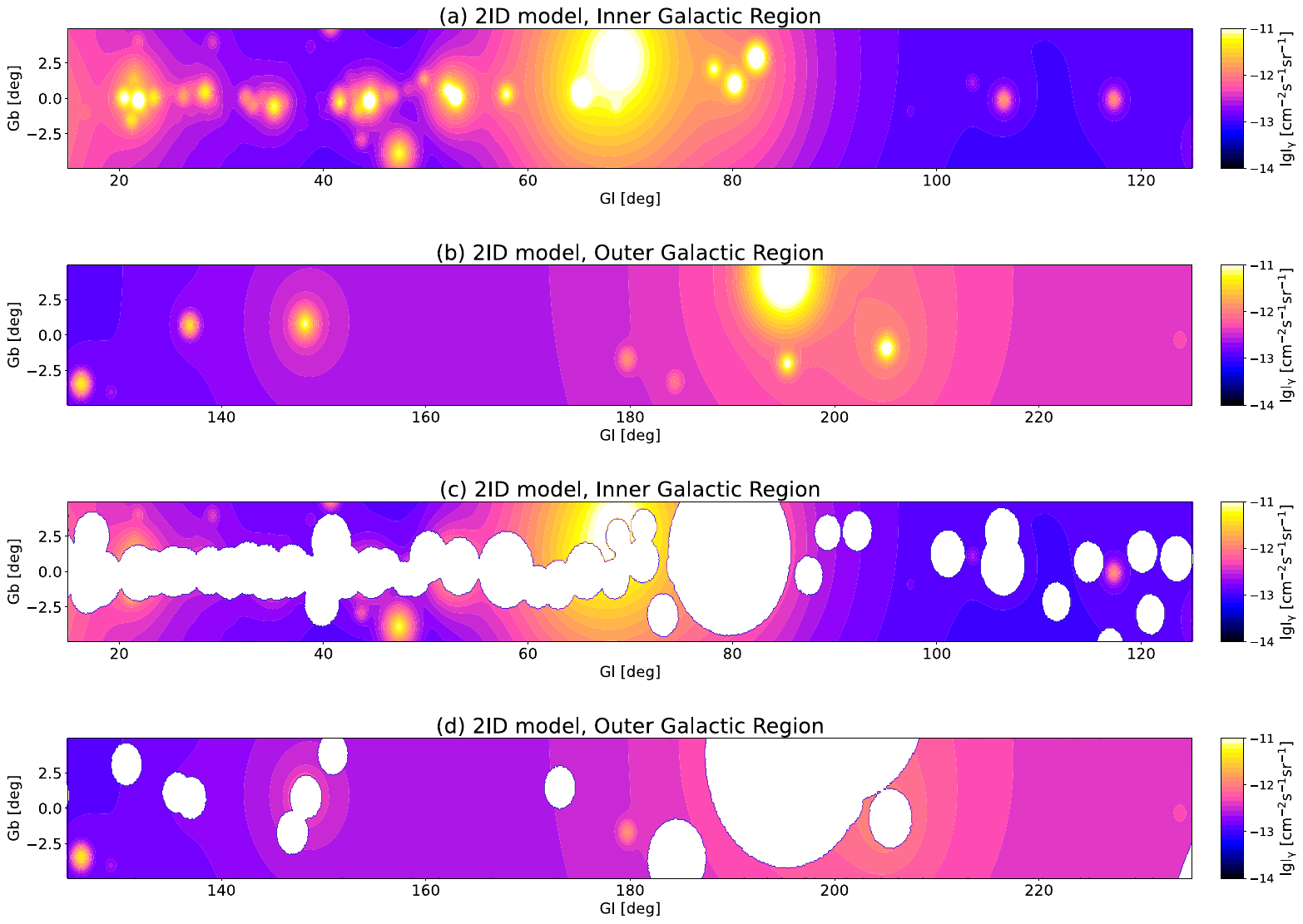}
\caption{Predicted gamma-ray intensity map in $25-100$\,TeV  contributed by halos of middle-aged pulsars recorded in ATNF catalog under the two-zone isotropic diffusion model (2ID model).
(a) the inner Galactic plane of $15^\circ<l<125^\circ$, (b) the outer Galactic plane of $125^\circ<l<235^\circ$. (c) the inner Galactic plane after masking the same region as removed in the LHAASO data analysis, (d) same as (c) but for the outer Galactic plane.}
\label{fig:hotmap0}
\end{figure*}

We then can simulate pulsar halos powered by each of selected pulsars based on the model. To compare with LHAASO's measurement, we make a two-dimensional intensity map of projection of each pulsar halo onto the Galactic plane according to the Galactic coordinate and distance of each pulsar, and then convolve the intensity map with LHAASO's point spread function \cite{LHAASO_whitepaper} to mimic the sky viewed by LHAASO. In Figure~3 and Figure~4, we present the projected intensity map of 1179 selected pulsars in the region where the DGE is extracted under the 2ID model and the AD model respectively. Then we mask the same region as used in the LHAASO data analysis, and calculate the average gamma-ray intensity in the remaining Galactic plane for $15^\circ<l<125^\circ$ and $125^\circ<l<235^\circ$ respectively. It is worth noting the existence of pulsars that are invisible to us, if their lighthouse-like radiation beam do not sweep Earth as they spin. These invisible off-beamed pulsars, however, may also generate pulsar halos around them and emission of these pulsar halos would contribute to DGE \cite{Yan23}. Therefore, the average gamma-ray intensity obtained above need be multiplied by a beam-correction factor following Ref.~\cite{Tauris98} to account for the contribution from halos of those off-beamed pulsars.

\begin{figure*}[htbp]
\includegraphics[width=\textwidth]{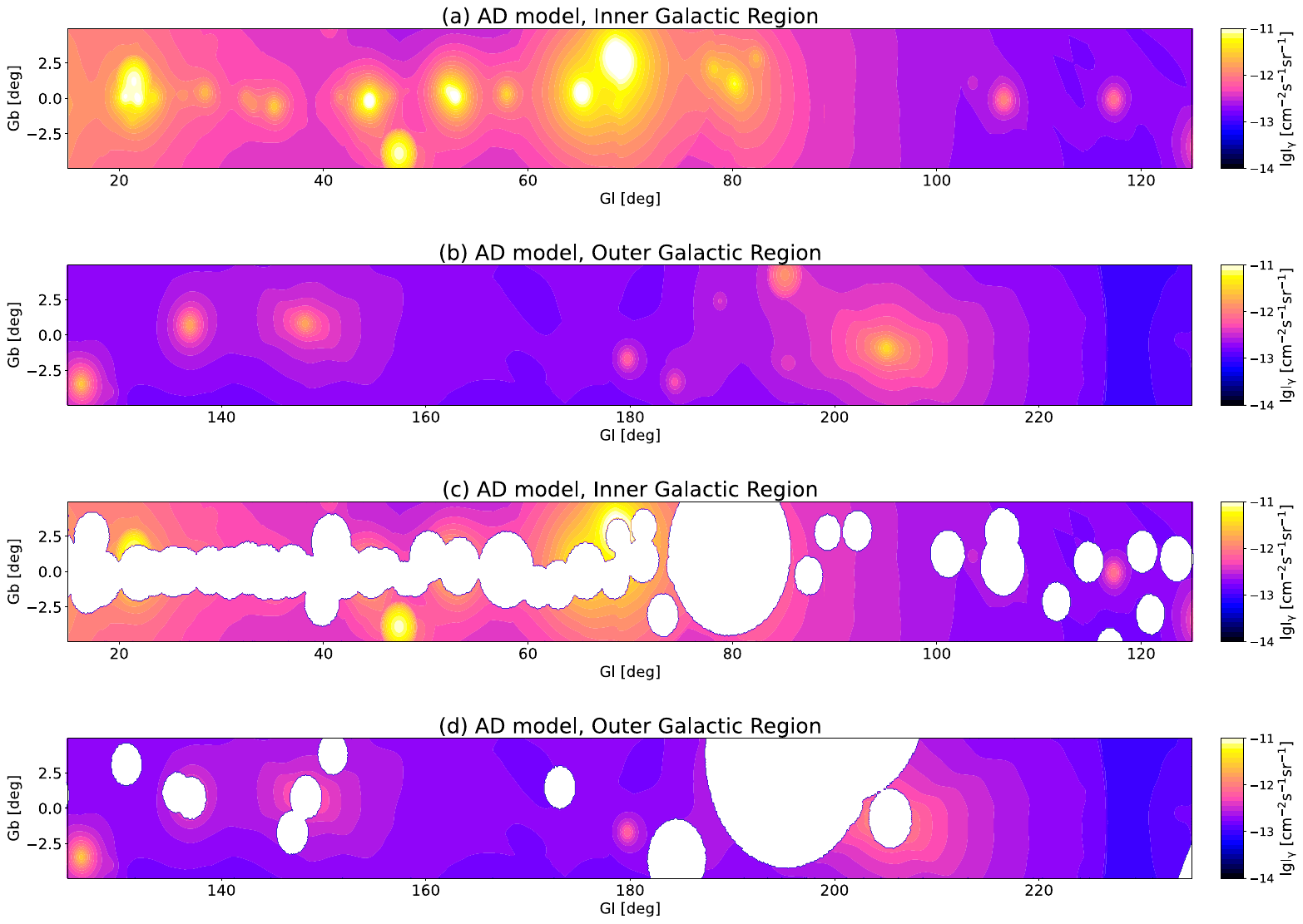}
\caption{ Predicted gamma-ray intensity map in $25-100$\,TeV contributed by halos of middle-aged pulsars recorded in ATNF catalog under the anisotropic diffusion model (AD model).
(a) the inner Galactic plane of $15^\circ<l<125^\circ$, (b) the outer Galactic plane of $125^\circ<l<235^\circ$. (c) the inner Galactic plane after masking the same region as removed in the LHAASO data analysis, (d) same as (c) but for the outer Galactic plane.}
\label{fig:hotmap1}
\end{figure*}

In Figure~5, we show that the DGE excess measured by LHAASO can be well explained by the pulsar halo population with a reasonable set of model parameters. Furthermore, an excess of DGE in $10-500\,$GeV is also reported in the same region of LHAASO's analysis based on Fermi-LAT's data \cite{ZhangR23}. Although the Fermi-LAT excess is not necessarily related to pulsar halos since there may be more potential GeV gamma-ray emitters in the Galaxy, we find that the Fermi-LAT excess can be accounted for by pulsar halos as well if some specific (but reasonable) model parameters are chosen. Beside the spectrum, we also compare the one-dimensional Galactic longitude profile between our simulation and LHAASO's observation. Because halos of those undetectable, off-beamed pulsars cannot be included in the modeled profile and because of the possible variation in the pair injection spectrum from individual pulsar halos, we do not expect a quantitative match between the measured profile and the modeled one. Nevertheless, as shown in Figure~6, we find the tendency of the longitudinal profile of the DGE excess and that predicted by our model are generally in good consistency, supporting the pulsar halo interpretation.

Our results suggest that pulsar halos make a considerable, if not dominant, contribution to the DGE emission measured by LHAASO below $\sim 60$\,TeV, extending probably down to $\sim 10\,$GeV. This conclusion does not depend on the model of pulsar halos and it only requires that approximately $10\%$ of a pulsar's spindown power has converted to electron/positron pairs escaping to ISM with an injection spectral index $s\gtrsim 2$. The main uncertainty lies in the halos powered by those off-beamed pulsars, which are now dealt with multiplying a beam-correction factor to contribution of halos formed around detected pulsars. If these invisible pulsars have similar properties to those detected ones in a statistical way (such as following the same luminosity function and spatial distribution), the total flux from the pulsar halo population predicted by the present model would be basically correct.

\begin{figure*}[htbp]
\centering
\includegraphics[width=\textwidth]{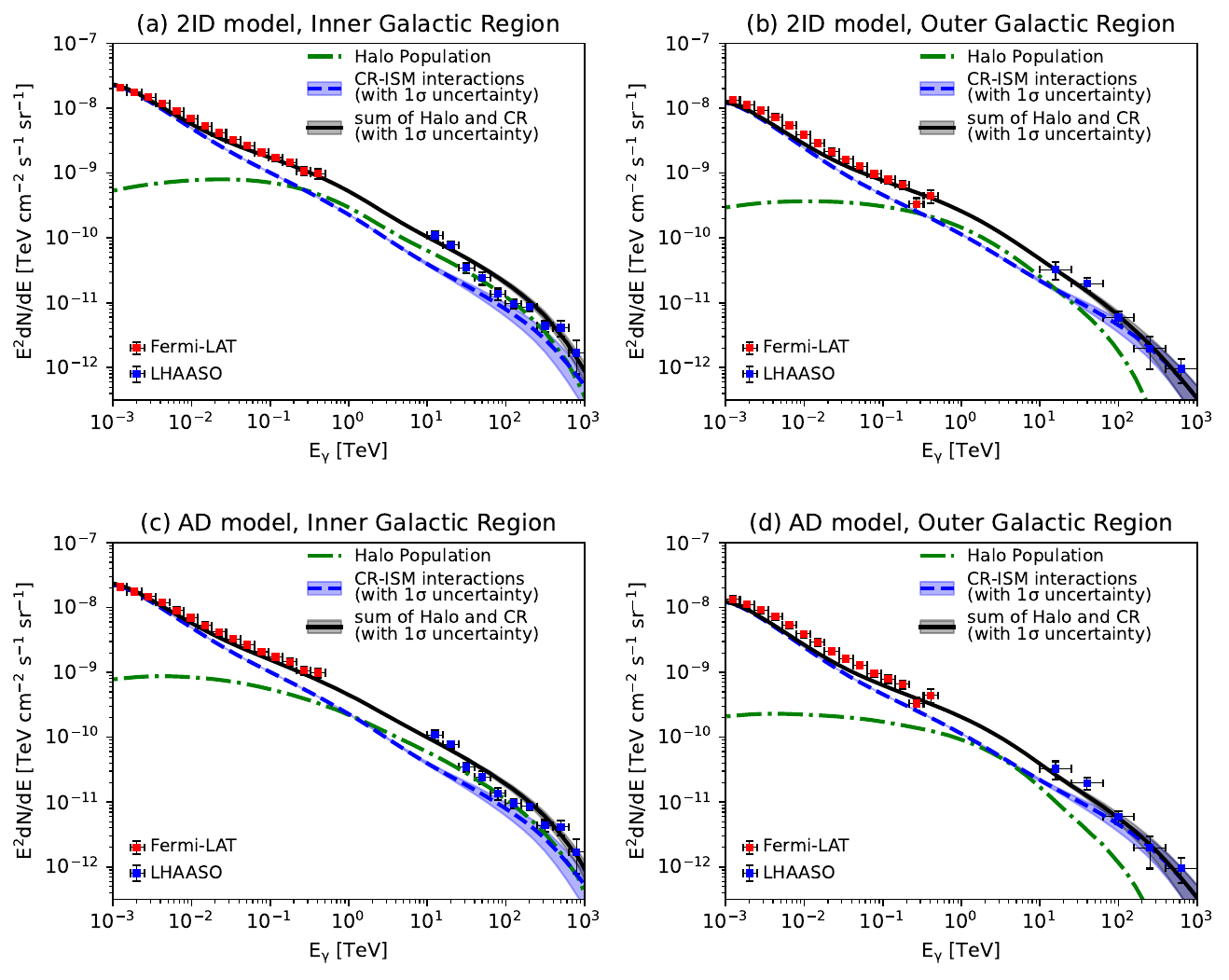}
\caption{Expected contribution of DGE from pulsar halos and CR-ISM interactions. (a) in the inner Galactic plane under the 2ID model. (b) in the outer Galactic plane under the 2ID model. (c) in the inner Galactic plane under the AD model. (d) in the outer Galactic plane under the AD model. In all four panels, blue dashed curves represent the gamma-ray flux generated by $pp$ collisions of CR hadrons (with the shaded region indicating the 1$\sigma$ uncertainty). Green dot-dashed curve represent the contribution of pulsar halos with $\rm \eta_e=0.15$ under the 2ID model and $\rm \eta_e=0.1$ under the AD model. In both models, the injection spectral index of pairs is $s=2.2$. Black curves show the sum of contributions of CR-ISM interactions and pulsar halos. Blue squares show LHAASO's measurement of DGE \cite{LHAASO_diffuse} while red squares represent the DGE measured by Fermi-LAT from the same region of LHAASO's analysis \cite{ZhangR23}.}
\label{fig:sed1}
\end{figure*}

The collective emission of pulsar halos constitutes a non-negligible background for other gamma-ray sources in the Milky Way. Therefore, it is important to understand the nature of these pulsar halos so as to better eliminate their influence when studying other sources. With increasing exposure time, sensitive TeV gamma-ray instruments such as LHAASO may resolve more and more pulsar halos, and hence reduces the background of this leptonic component. However, it would not be straightforward to recognize halos of those off-beamed pulsars. Accurate measurements of the intensity profile of gamma-ray sources by imaging air Cherenkov telescopes of high angular resolution such as the Cherenkov Telescope Array \cite{CTA11}, as well as a multiwavelength study, would help the identification. On the other hand, the neutrino measurement is crucial to distinguish the hadronic component from the leptonic component. The next-generation neutrino telescopes with improved angular resolution and effective area would be capable of identifying bright neutrino sources from the Galaxy. The advanced measurement of high-energy neutrinos, along with the high-sensitivity gamma-ray observations, will hopefully unravel the composition of the DGE, uncover the nature of Galactic leptonic and hadronic gamma-ray emitters, and finally solve the century-old puzzle of the cosmic-ray origin.

\begin{figure*}[htbp]
\includegraphics[width=\textwidth]{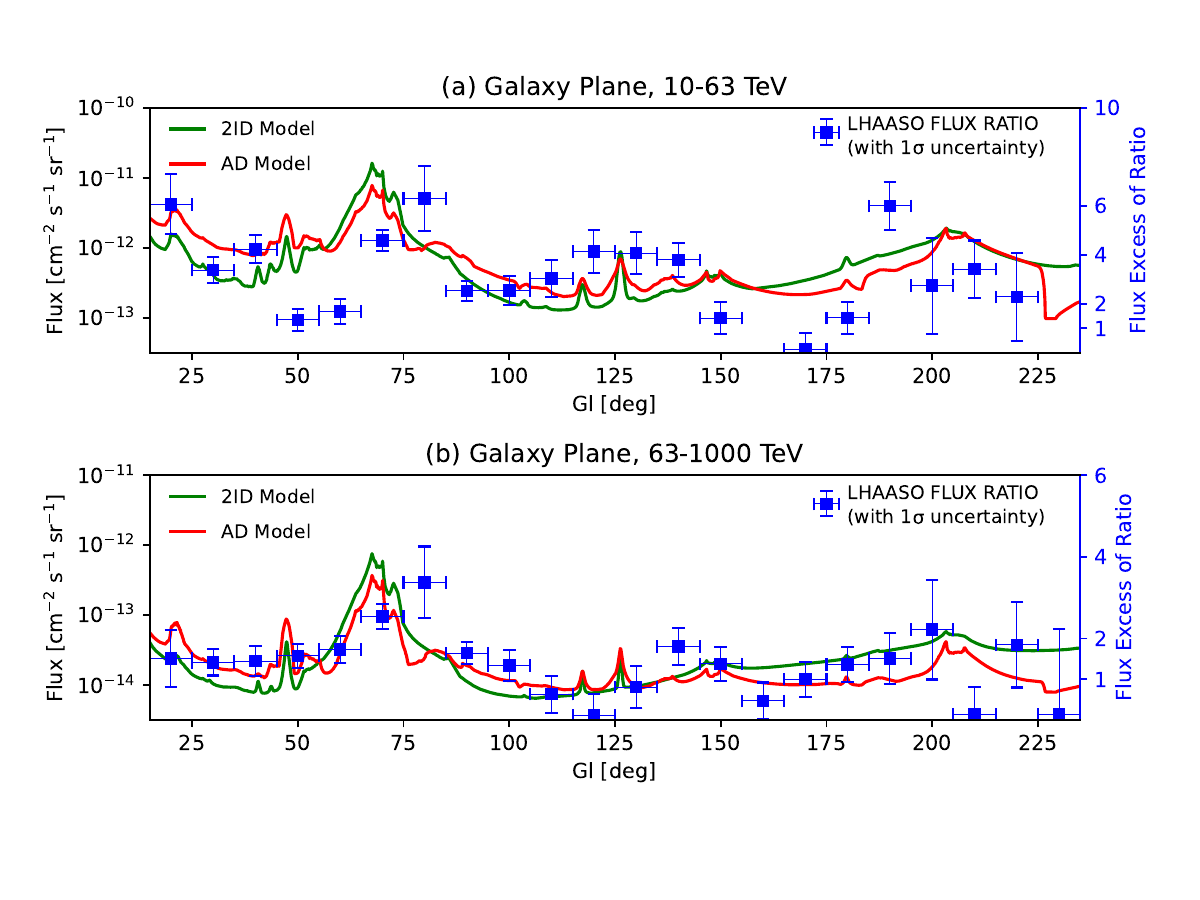}\\
\caption{ Galactic longitudinal gamma-ray profile contributed by pulsar halos. (a) in the energy range of $10-63$\,TeV. (b) in the energy range of $63-1000$\,TeV.
In both panels, green lines and red lines represent cases of the 2ID model and the AD model, respectively.
Also shown in the flux excessive ratio of LHAASO's measurement to the predicted one from CR-ISM interactions by blue squares (error bars indicating 1$\sigma$ uncertainties).}
\label{fig:profile}
\end{figure*}


\section*{Method}
\section{Source Contribution to the Neutrino flux}
The neutrino emission measured by IceCube contains a considerable contribution from Galactic neutrino emission sources. To estimate the source component in the measured neutrino flux and the converted hadronic gamma-ray flux, we add up the gamma-ray  sources of potential hadronic origin observed by HESS and LHAASO. Among the 90 sources listed in the 1st LHAASO catalog, 85 of them are Galactic sources. Among the 85 Galactic sources, we exclude 35 sources with pulsar or PWN associations which are like of the leptonic origin, and focus on those sources located within Galactic latitude $|b| \le 5^{\circ}$. There are 45 remaining gamma-ray sources of potential hadronic origin in the Galactic plane, measured in the range of 1-25$\rm \,TeV$ by LHAASO-WCDA and $\rm \ge 25\,TeV$ by LHAASO-KM2A. Note some sources only have detection by only one of the two detectors of LHAASO. The best-fit spectral parameters and the uncertainties are given in Ref.\cite{LHAASO_catalog}. On the other hand, HESS Galactic Plane Survey (HGPS) \cite{HGPS_2018} announced 78 sources from 300\,GeV to 30\,TeV. Among them, 14 are confirmed PWNe and others are regarded as potential hadronic sources. Part of them are the same sources recorded in the LHAASO catalog so we only consider 43 sources in HGPS with declination smaller than $-20^\circ$, where is outside LHAASO's field of view.

Both catalogs provide the best-fit spectra of these sources either described with a power-law function
\begin{equation}
    F(E) = N_0 (\frac{E}{E_0})^{-\Gamma},
\end{equation}
or exponential cutoff power-law function from
\begin{equation}
    F(E) = N_0 (\frac{E}{E_0})^{-\Gamma} \exp(-\frac{E}{E_{\rm c}}),
\end{equation}
as well as the uncertainties of these parameters, i.e., $\Delta N_0$, $\Delta \Gamma$, and $\Delta E_{\rm c}$. The statistic uncertainty of flux is evaluated via
\begin{equation}
    \Delta F_{stat}(E) = F(E)\sqrt{\frac{\Delta N_0^2}{{N_0}^2} + \log^2(\frac{E}{E_0})\Delta \Gamma^2 +
    \frac{E^2 \Delta E_{\rm c}^2}{{E_{\rm c}}^4}
    }
\end{equation}
noting that the last term in the square root should be dropped out if the spectrum is fitted with simply a power-law function without cutoff. Apart from that, we also calculate the systematic uncertainty. In the case of WCDA data, the systematic uncertainty is estimated to be $^{+8\%}_{-5\%}$ on the flux and 0.02 on the spectral index. In the case of KM2A data, the systematic uncertainty is estimated to be $7 \% $ on the flux and 0.02 on the spectral index. And for HESS data, the systematic uncertainty is estimated to be $30 \% $ on the flux and 0.2 on the spectral index. The total statistic flux is added up via $F_{\rm tot}(E)=\sum\limits_i F_i(E)$ where $F_i$ represents the spectrum of the $i$th source in the list, and the statistic uncertainty is calculated by $\Delta F_{\rm tot,stat}(E)=\sqrt{\sum\limits_i \Delta F_{i,stat}(E)^2}$. The total systematic uncertainty is calculated by $\Delta F_{\rm tot,sys}(E)=\sum\limits_i \Delta F_{i,sys}(E) $. And the total uncertainty is evaluated by $\Delta F_{\rm tot}(E)=\sqrt{\Delta F_{\rm tot,stat}(E)^2+\Delta F_{\rm tot,sys}(E)^2}$.

On the other hand, the neutrino flux is obtained based on certain all-sky template for hadronic gamma-ray emission in IceCube's analysis \cite{IceCube23}. To compare with the hadronic emission from Galactic sources as discussed before, we exclude the region outside $|b| \le 5^{\circ}$. The integrated flux of the remaining region is 50.2\% and 71.9\% of the all-sky integrated flux in $10\,$TeV for the Fermi-LAT $\pi^0$ template and the KRA$_\gamma^5$ template respectively. The inferred gamma-ray fluxes in the ROI of $|b| \le 5^{\circ}$ derived from the neutrino measurement and that from hadronic sources are compared as shown in Figure~1.

To obtain the DGE of the hadronic origin from the same region of LHAASO's measurement, we need to further exclude the region outside $15^\circ<l<235^\circ$ and $|b|<5^\circ$ in both the $\pi^0$ template and the KRA$_\gamma^5$ template. In addition, we also need to mask the same region as that in LHAASO's DGE analysis \cite{LHAASO_diffuse}. The remaining region is shown in  Extended Data Figure~1. For the inner (outer) Galactic region $15^\circ<l<125^\circ$ ($125^\circ<l<235^\circ$), the integrated flux of the remaining region is 6.87\% (5.69\%) and 5.74\% (2.49\%) of the all-sky integrated flux in $10\,$TeV for the Fermi-LAT $\pi^0$ template and the KRA$_\gamma^5$ template respectively. Note that the KRA$_\gamma^5$ template is mildly energy dependent, and consequently the fraction of the remaining flux also changes with energy slowly for this template. This effect is considered in our analysis and the dependence of the fraction as a function of energy is shown in Supplemental Material. The corresponding solid angle of the remaining region is 0.206\,sr for the inner Galactic plane and 0.268\,sr for the outer Galactic plane. The neutrino flux of single flavor (assuming a flavor ratio of $1:1:1$ after oscillation) is related to the co-produced gamma-ray flux of the hadronuclear origin by $E_\gamma^2\frac{dN_\gamma}{dE_\gamma}=2E_\nu^2\frac{dN_\nu}{dE_\nu}$,
where $E_\gamma=2E_\nu$.
The inferred DGE of the hadronic origin can be obtained via dividing the flux by the solid angle.  We note that our result for the total source contribution shown in Figure~1 is consistent with an independent study \cite{Fangke23} appearing online at the finalizing stage of this study.

\renewcommand{\figurename}{Extended Data Figure}
\setcounter{figure}{0}

\begin{figure*}[htbp]
\includegraphics[width=\textwidth]{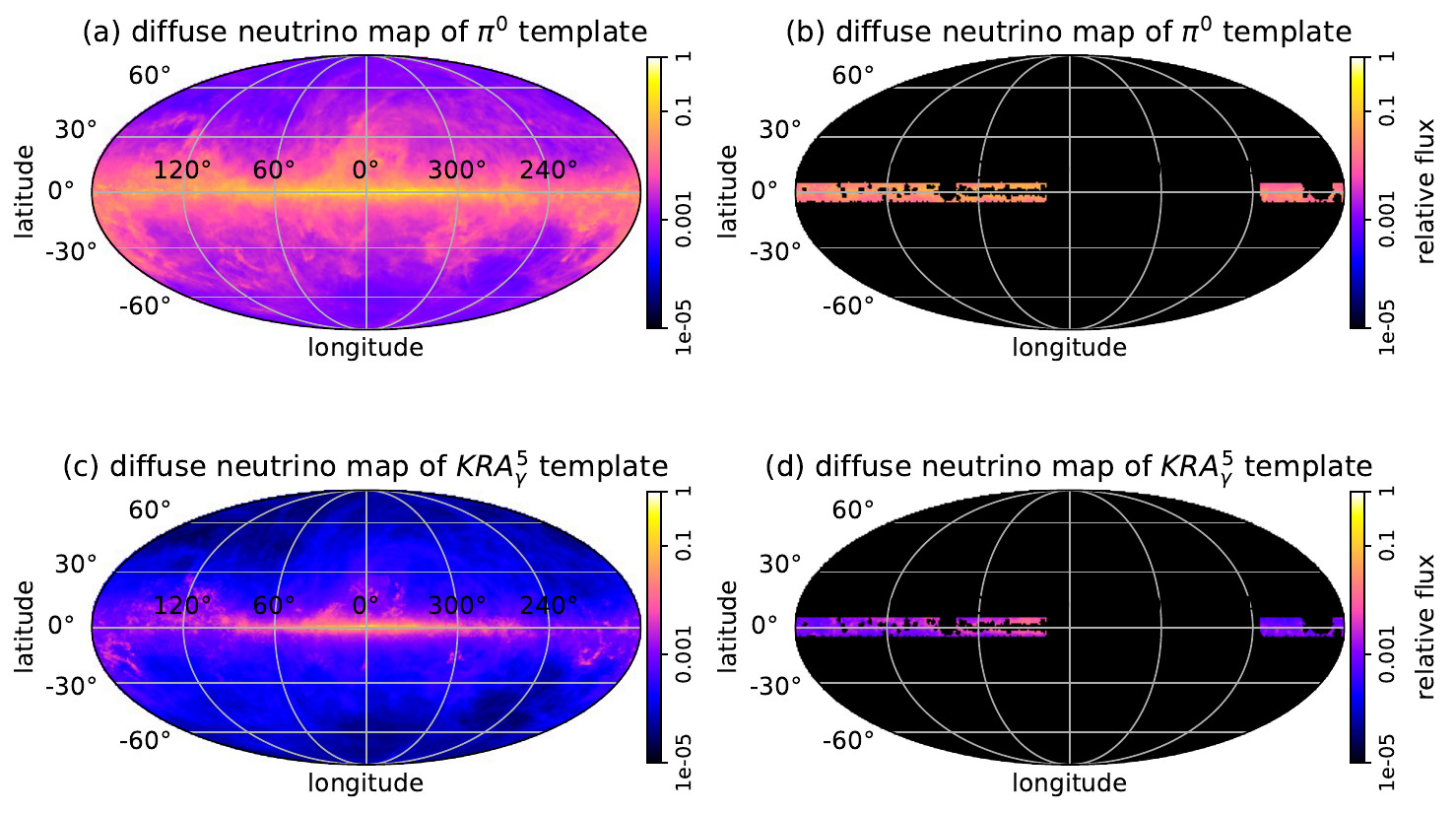}\\

\caption{Neutrino intensity map with the $\pi^0$ template and KRA$_{\gamma}^{5}$ template. (a) the all-sky intensity map with the $\pi^0$ template. (b) the all-sky intensity map with the KRA$_{\gamma}^{5}$ template. (c) the intensity map in the same ROI of LHAASO's analysis, after removing the region out of the Galactic plane and masking the LHAASO source region, for the $\pi^0$ template. (d) same as (c) for the KRA$_{\gamma}^{5}$ template.}

\label{fig:ic_map}
\end{figure*}

\section{Injection of Electron/Positron Pairs}
We assume that pairs are injected in a power-law spectrum with a high-energy cutoff at a rate of
\begin{equation}
Q\left(E_{\rm e}, t\right)=Q_{\rm 0}\left(t \right) E_{\rm e}^{-s} e^{-E_{\rm e} / E_{\rm \max }}
\label{eq:Qe}
\end{equation}
where $Q_0$ is the normalization which is related to the spindown power of the pulsar $L_{\rm s}$ by
\begin{equation}
\int E_{\rm e}Q(E_{\rm e},t)dE_e=\eta_e L_{\rm s}(t).
\end{equation}
The temporal evolution of the spindown power can be given by $L_{\rm s}(t)=\eta_e L_{\rm s, 0}/(1+t/\tau_0)^2$ \cite{Pacini73}, simply assuming pulsars as a rotational magnetic dipole. The present spindown power can be obtained by $L_{\rm s}=-4\pi^2I\dot{P}P^{-3}$ once the rotational period $P$ and the first derivative of the period $\dot{P}$ of the pulsar is measured, with $I$ being the pulsar's moment of inertia which is taken to be a typical value of $10^{45} \, \rm g ~cm^2$. Here $L_{\rm s, 0}$ is the initial spindown power of the pulsar and $\tau_0$ is the initial spindown timescale which can be derived given $P$, $\dot{P}$ and the initial rotational period $P_0$. The initial rotational period of a pulsar is an unknown parameter, we here assume it to be
\begin{equation}
    P_0=\left\{
\begin{aligned}
& 0.1P, \  P > 300\,\rm ms ,\\
& 30\, {\rm ms},  \ P \le 300\,\rm ms.
\end{aligned}
\right.
\end{equation}
The cutoff energy $E_{\rm max}$ in the injection spectrum describes the acceleration limit of the PWN and can be given by $E_{\rm max}=(2\eta_{\rm B})^{1/2}e(L_{\rm s}/c)^{1/2}$ \cite{deJager92}. $\eta_{\rm B}$ is the fractional energy converted from kinetic energy of the pulsar wind to the magnetic energy at the pulsar wind termination shock, which is suggested to be $\gtrsim 0.1$ for middle-aged pulsar \cite{Yan23}. For simplicity, we take $\eta_{\rm B}=0.1$ here for all pulsars. The beam-correction factor $f_{\rm beam}$ depends on the age or rotational period of the pulsar, which reads \cite{Tauris98}
$f_{\rm beam} = 0.011\left[\rm log\left( \tau_{\rm age}/100{\rm Myr}\right) \right]^{2}+0.15$ where $\tau_{\rm age}$ is the age of the pulsar, or $f_{\rm beam}=0.09[\log(P/10{\rm s})]^2+0.03$. Both relations yield a similar beam-correction factor. The flux of each halo is divided by the corresponding beam-correction factor to account for the contribution of those off-beamed pulsar with similar properties.

\section{Models for Pulsar Halos}
Pulsar halos are extended gamma-ray sources. Therefore, even if a pulsar is located inside the masked region of the LHAASO's analysis, the halo may extend beyond the region and the outer part of its emission would be counted into the DGE. Similarly, even if a pulsar is located outside the masked region, part of its halo may still overlap with the masked region and the emission is removed. Therefore, we need to model the spatial morphology of each halo, so that we can quantitatively calulate their contribution to DGE by masking the same area of the sky as did LHAASO's analysis. In this study, we consider two models for pulsar halos. Note that we do not model the emission of their corresponding PWNe  which are supposed to present in central part of the halos. PWNe of the Geminga pulsar and the Monogem pulsar are not accompanied with apparent TeV emission. Also, a search of TeV emission by VERITAS around three bow-shock PWNe (powered by middle-aged pulsars) result in null detection\cite{Benbow21}. These observational facts imply that TeV emission from PWNe of middle-aged pulsars are unimportant.

\noindent {\bf Two-zone Isotropic Diffusion (2ID) model}\\
Under this model , the diffusion coefficient is suppressed in the inner zone to a radius of $r_{\rm b}$ (typically a few tens of parsecs), and beyond the suppressed diffusion zone, a typical diffusion coefficient is assumed, i.e.,
\begin{equation}
D(E,r)=\left\{
\begin{aligned}
& D_0(E/100 \, {\rm TeV})^{1/3}, \  r < r_b ,\\
& D_{\rm ISM}(E/100 \, {\rm TeV})^{1/3},  \ r \ge r_b.
\end{aligned}
\right.
\end{equation}
For simplicity we fix $r_b=20 \, \rm pc$, $D_0=4.5\times 10^{27} \, \rm cm^2 s^{-1}$, $D_{\rm ISM}=1.8\times 10^{30} \, \rm cm^2 s^{-1}$ \cite{Trotta11}. The analytical solution of particle distribution in both momentum space and real space in the 2ID model is given by Ref.\cite{Tang19}. For cooling of pairs, we consider synchrotron cooling in the Galactic magnetic field, and the IC cooling in the background radiation field including CMB and ISRF from the far infrared band to the ultraviolet band. The interstellar magnetic field strength around each pulsar is based on the model proposed by Jansson \& Farrar \cite{JF12_regular, JF12_random}, as shown in Extended Data Figure~2. The Monte Carlo method is used to generate the magnetic field strength. The ISRF is based on the model proposed by Popescu et al.\cite{Popescu17}.

\vspace{5pt}
\noindent {\bf Anisotropic Diffusion (AD) model}\\
Under the AD model, the perpendicular diffusion coefficient $D_{\perp}$ is suppressed by a factor of $M_{\rm A}^4$ with respect to the parallel diffusion coefficient $D_{\parallel}$, i.e., $D_\perp=M_A^4D_\parallel$. We assume $D_\parallel$ to be the typical diffusion coefficient in the ISM, i.e., same as $D_{\rm ISM}$ in the 2ID model. We follow the analytical solution provided by Refs.~\cite{Delahaye10} to calculate the distribution of pairs in both real space and energy space. The same GMF and ISRF are considered in this model to calculate the cooling of pairs. Note that injected pairs diffuse preferentially along the mean magnetic field direction in the AD model, and hence the magnetic field direction matters. Therefore, in addition to the magnetic field strength, we also need to consider the magnetic field direction in the AD model. The GMF in the JF12 model contains both a regular component and a random component. For the regular component, we can simply obtain the direction of the regular field around each pulsar based on the positions of the pulsar in the Milky Way. For the random striated component, the so called ``ordered random fields'', is randomly assigned either a direction aligned with the regular component or anti-aligned with the regular component \cite{JF12_regular}. For those ``true random fields'', which may represent the influence of local astrophysical objects, we assign a random direction with an equal possibility in the $4\pi$ solid angle. The final direction (as well as the strength) of the magnetic field is determined by the vector sum of the regular component and the random component.

\begin{figure*}[htbp]
\includegraphics[width=0.5\textwidth]{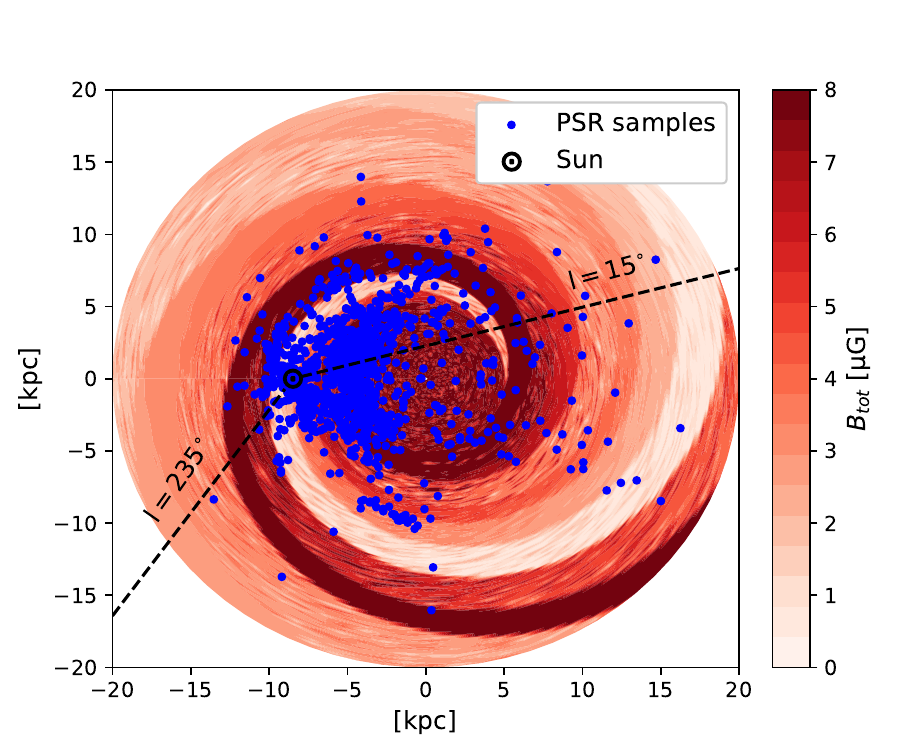}\\
\caption{Distribution of the strength of the magnetic field in the Galactic plane. The Galactic center is located at (0,0). The positions of pulsars considered in this work are marked as blue dots. The boundary of the Galactic longitude of LHAASO's ROI $l=15^{\circ}$ and $l=235^{\circ}$ are labeled for as dashed line for reference.}
\label{fig:Btot}
\end{figure*}

After obtaining the pair distribution around each pulsar, we can calculate the gamma-ray emissivity of pairs through the IC process. By integrating the emission over the line of sight (LOS) toward a certain direction in the Galactic plane, we can project each gamma-ray halo onto the plane of the sky, and obtain the 2D intensity map composed of emission from pulsar halos as shown in Figure~3 and Figure~4. A detailed description of the procedure and relevant discussion is provided in Supplemental Material.

\vspace{5pt}

\section*{Acknowledgement}
We would like to thank referees for their constructive comments and suggestions. We are also grateful to Lu Lu for the helpful discussion on IceCube's results and Ke Fang for the valuable comment.
This work is supported by the National Natural Science Foundation of China (No. U2031105, 12393852, 12220101003, 12333006) and the Project for Young Scientists in Basic Research of Chinese Academy of Sciences (No. YSBR-061).

\clearpage

\setcounter{section}{0}


\section*{Supplemental Material}
\section{Sources below HGPS's sensitivity}
\noindent We note that the source list considered is not complete due to the limitation of the instrument's sensitivity, and therefore we may underestimate the source contribution to the Galactic neutrino emission. The underestimation is more  significant in the southern sky, because the sensitivity of HGPS is lower than that of LHAASO above 10\,TeV. Assuming that sources in the southern sky has the same properties with sources in the northern sky in a statistical way, we may estimate how much the total source flux would increase if HGPS could reach the same sensitivity of LHAASO. To do this, we check the detectability of LHAASO sources by HESS with the same exposure in HGPS. The sensitivity of HGPS reaches about 1.5\% of Crab flux\footnote{Crab flux is defined as $F(1-30\, \rm {TeV})=\rm 2.26\times 10^{-11} \, cm^{-2}s^{-1}$} for point-like source \cite{HGPS_2018}. Its sensitivity for extended sources measured by LHAASO can be estimated via multiplying the HGPS point-source sensitivity by a factor of $(1.51/0.08)r_{39}$, where $r_{39}$ is the 39\% containment radius of an extended source (in degree) measured by LHAASO, $1.51r_{39}$ gives the 68\% containment radius of the source, and 0.08 is the 68\% containment radius of the point-spread function of HESS (in degree). If the integrated flux of the LHAASO source in $1-30$\,TeV is below the HGPS extended-source sensitivity, we regard the source undetectable in HGPS. After performing the examination for each source recorded in the 1st LHAASO catalog in the Galactic plane, we find that the total flux of detectable sources in HGPS is about 33.4\% of the total flux of all the examined sources. Therefore, we may arrive at that the total HGPS source flux would increase by about 50\% if HGPS has the same sensitivity of LHAASO. Consequently, the fraction of the remaining neutrino flux (based on the $\pi^0$ template) after subtracting hadronic source contribution would decrease from $\sim$1/3 down to $\sim$1/4 of the total flux reported by IceCube. Considering the contribution of unresolved sources would result in a more significant discrepancy between the expected DGE flux of hadronic origin and the observed flux. This would leave a larger room for an additional component of leptonic origin in the DGE.

\section{2D Gamma-ray Intensity Map}
After obtaining the pair distribution around each pulsar, we can calculate the gamma-ray emissivity of pairs through the IC process. By integrating the emission over the line of sight (LOS) toward a certain direction in the Galactic plane, we can project each gamma-ray halo onto the plane of the sky, and obtain the 2D intensity map composed of emission from pulsar halos. For the 2ID model, we follow the projection method shown in Ref.~\cite{Liu19} to overlaid each pulsar halo in the Galactic plane. For the AD model, the projection method is detailed Ref.~\cite{Liu19_prl} but additional procedure is carried out to minimize the possible bias arising from the randomly generated magnetic field direction, as the latter largely determines the morphology of pulsar halos in the AD model. To do this, we generate 10,000 realizations of the GMF, and obtain intensity maps in the Galactic plane composed of pulsar halos based on each of the realization of the GMF.

\noindent For reference, we show the 2D probability distribution of  the magnetic field direction with respect to the cosine of the angle between the generated field and of the LOS ($\cos \phi$), and the azimuthal angle of projected field on the celestial plane ($\zeta$) in Supplementary Fig.~\ref{fig:B2}. In left panels, we show a pulsar located in the magnetic field dominated by the random component. In this case, the probability distribution of $\cos\phi$ and $\zeta$ is almost homogeneous (the top-left panel) and the average expectation of the halo morphology is nearly isotropic (the bottom-left panel). On the contrary, if a pulsar located in the magnetic field dominated by the regular component, as shown in the right panels, the average magnetic field would have a preferential direction. This can be seen from the enhancement of the probability in the lower-middle part of the $\cos\phi - \zeta$ plane. The resulting halo is correspondingly elongated along the projected direction of the magnetic field in the celestial plane. 

\noindent Finally, we average over all the generated gamma-ray intensity maps of the Galactic plane, and get the average expectation of the DGE contributed by pulsar halos under the AD model. Particularly, for the observed two pulsar halos around Geminga and Monogem, we restrict the inclination angle between the magnetic field and the LOS to be less than $5^{\circ}$, in order to be consistent with the observed morphology of these two halos \cite{Liu19_prl}.

\renewcommand{\figurename}{Supplementary Figure}
\setcounter{figure}{0}

\begin{figure*}[htbp]
\includegraphics[width=0.4\textwidth]{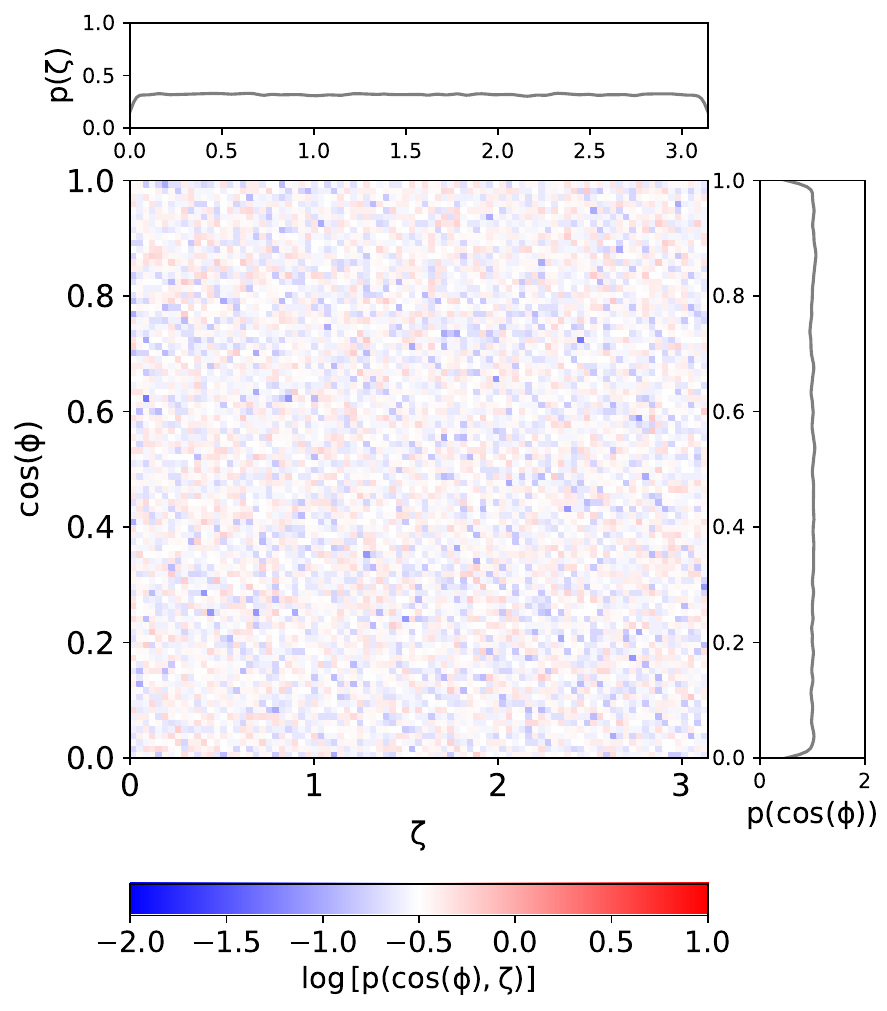}
\includegraphics[width=0.4\textwidth]{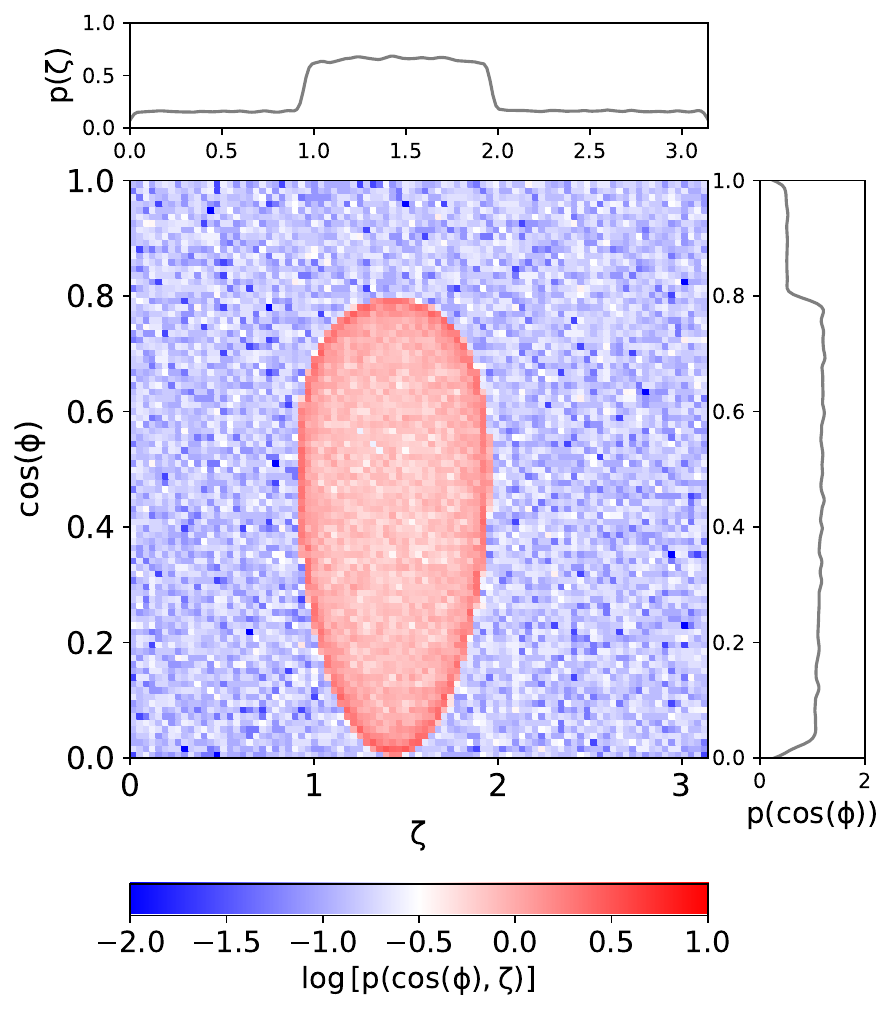}\\
\includegraphics[width=0.4\textwidth]{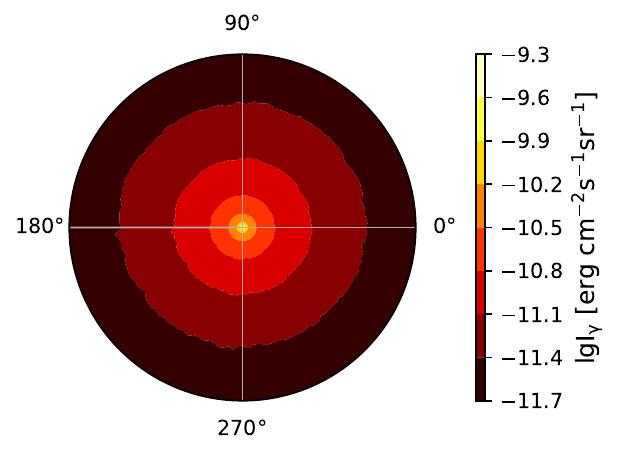}
\includegraphics[width=0.4\textwidth]{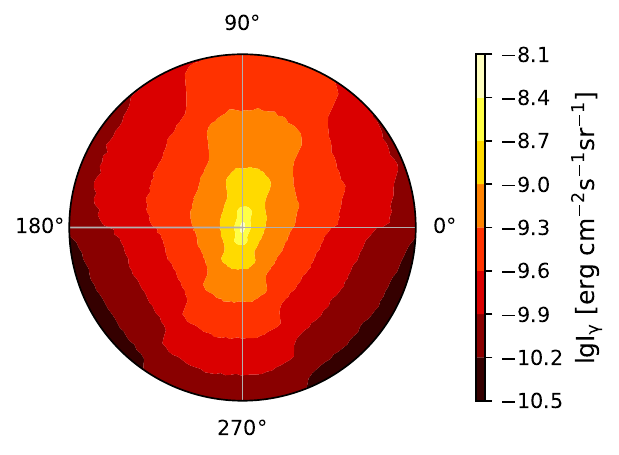}\\
\caption{2D probability density distribution of the magnetic field geometric parameters around two example pulsars and the resulting pulsar halo morphology.  Magnetic field orientation parameters are the cosine of the angle between the magnetic field and the LOS ($\cos \phi$), and the azimuthal angle of projected field on the celestial plane ($\zeta$). In the left panel, the pulsar is located in the magnetic field dominated by the random component, and thus the 2D probability distribution of the magnetic field distribution is approximately uniform and the mean morphology of the resulting halo is nearly isotropic. In the right panel, the pulsar is located in the magnetic field dominated by regular component, and thus the 2D probability distribution of the magnetic field distribution tends to cluster around the direction of the regular magnetic field, and the resulting halo is elongated along the projected direction of the magnetic field in the celestial plane. For both two cases, the halo morphology is averaged over 10,000 realizations of the generated GMF.}
\label{fig:B2}
\end{figure*}

\section{Influences of the age cut criterion of pulsars and the initial rotational period $P_0$}

\noindent We select pulsar samples with characteristic age between 50 kyr and 10 Myr. This age cut criterion is to exclude the contribution of relatively young pulsars and millisecond pulsars. For those relatively young pulsars, the injected electrons are still well confined in their PWNe and do not form pulsar halos. PWNe around those relatively young pulsars can contribute to the diffuse gamma-ray emission in principle \cite{20Cataldo}.
However, in this work those pulsars with bright PWNe have been removed through the mask procedure. In Supplementary Fig.~\ref{fig:young_psr}, we show that among 91 pulsars with characteristic age $\tau_c \rm < 50 \, kyr$, only two pulsars (PSR J0501+4516 and PSR J0729-1448) are out of the mask region. PSR J0501+4516 has a very low spin-down luminosity ($1.2 \times 10^{33} \, \rm erg \, s^{-1}$) and could not contribute significantly to the DGE. PSR J0729-1448 is of a higher spin-down luminosity ($2.8 \times 10^{35} \, \rm erg \, s^{-1}$) and has relatively small distance from the Earth (2.68\,kpc). This pulsar is associated with an X-ray PWN \cite{2008Chandra} but has no gamma-ray PWN detection yet, which might result from a high magnetic field suppressing the IC emission. Whether PSR J0729-1448 has an undetected faint gamma-ray PWN or not, its contribution to the DGE over the ROI is negligible. For millisecond pulsars, the ambient environment is complex and we ignore their possible contribution on the conservative side, although they could power pulsar halos as well\cite{Hooper22}. While the exact value of the age cut is uncertain, we show that, within reasonable ranges suggested by ages of those observed pulsar halos, the variations of the age cuts have no great influence (within a factor of 1.5) on the predicted gamma-ray flux in the energy range of LHAASO's measurement (i.e., above 10\,TeV), as illustrated in the upper panels of Supplementary Fig.~\ref{fig:factor_spec}.

\noindent The value of $P_0$ can affect the initial spindown luminosity of a pulsar and hence largely determines the particle injection rate at early time. Given the sum of the magnetic field energy density and the radiation field energy density in the ISM to be $\sim 10^{-12}\,\rm erg/cm^3$ typically, electrons diffusing in the ISM will cool within a timescale
\begin{equation}
    t_{c}\simeq 16 \left(\frac{E_e}{30\, \rm TeV}\right)^{-1}\left(\frac{U_B+U_{\rm rad}}{10^{-12}\rm erg~cm^{-3}}\right)^{-1}\, {\rm kyr}.
\end{equation}
Electrons emitting $>10$\,TeV photons generally have energies above 30\,TeV. As a result, we would expect electrons accounting for $>10\,$TeV gamma rays injected at recently and hence are not related to the injection at early time, although the electrons of energy below 10\,TeV could be subject to the assumed value of $P_0$. As illustrated in the lower panels of Fig.~\ref{fig:factor_spec}, we show the influence of the initial period on the predicted DGE flux from pulsar halos. We see a strong dependency of the flux on $P_0$ below several TeV, but the flux is almost unaffected above 10\,TeV. Due to the same reason, a different braking index does not significantly affect the obtained flux above 10\,TeV neither.

\begin{figure*}[htbp]
\includegraphics[width=\textwidth]{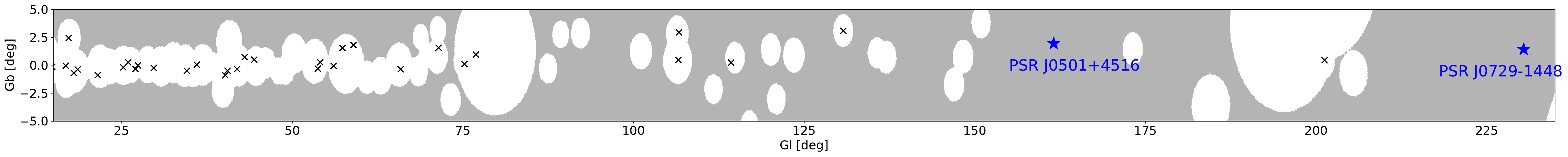}\\
\caption{Distribution of pulsars with characteristic age $\tau_{\rm c} \rm < 50 \, kyr$ in the Galactic plane. White region represent the masked sky in LHAASO's DGE analysis. Black crosses representing pulsars within the masked region and two blue stars show positions of PSR~J0501+4516 and PSR~J0729-1448 which are outside the masked region.}
\label{fig:young_psr}
\end{figure*}

\begin{figure*}[htbp]
\centering
\includegraphics[width=\textwidth]{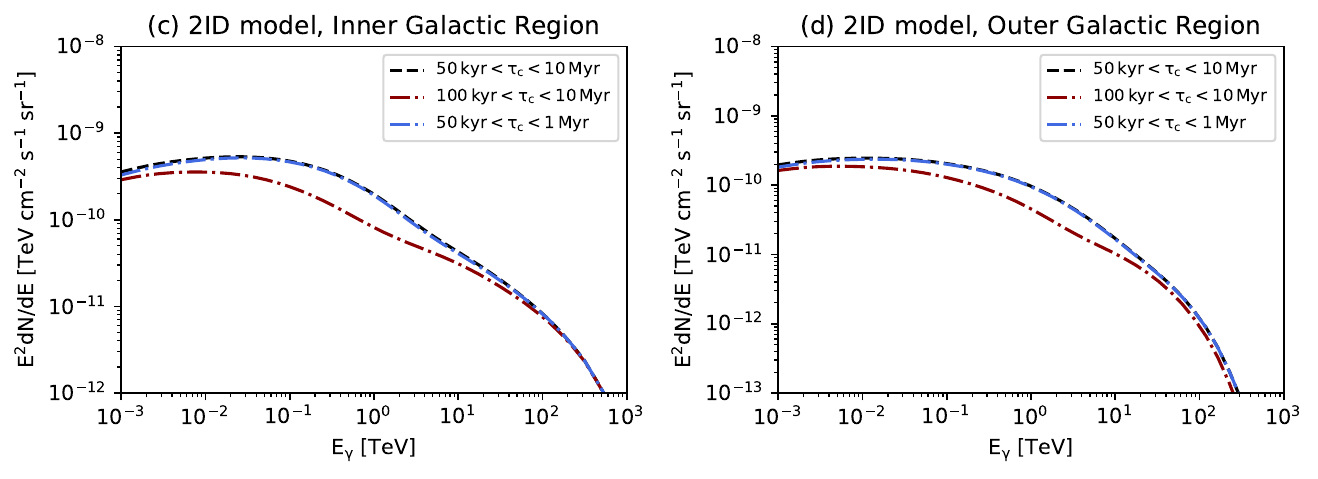}\\
\includegraphics[width=\textwidth]{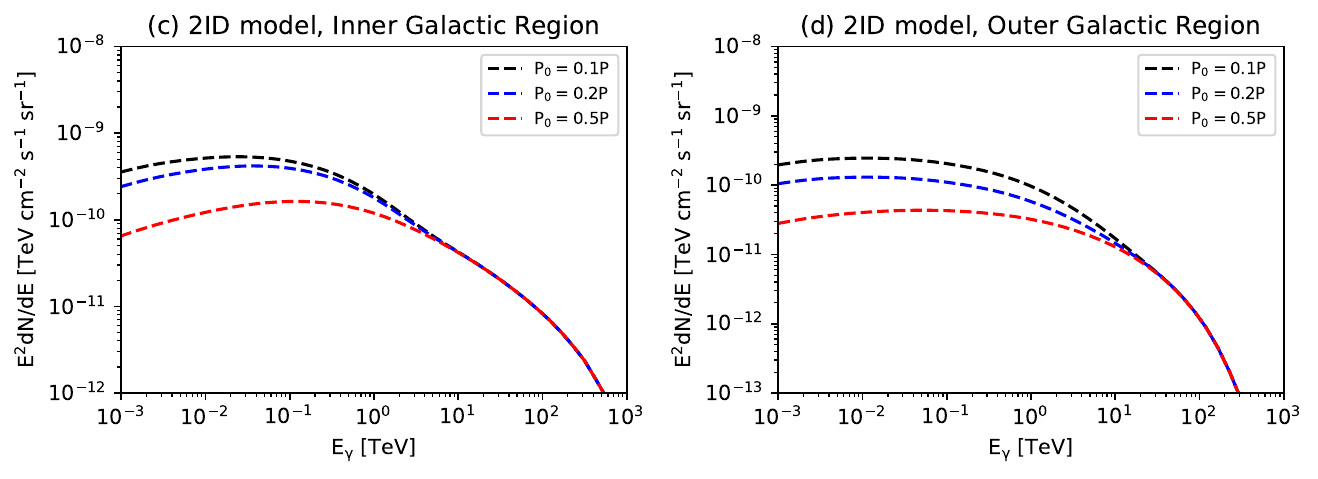}
\caption{Spectral energy distribution (SED) of the population of unresolved pulsar halos. The left panels showcase the inner Galactic Region, while the right panels showcase the outer Galactic Region. The upper panels show the influence of the age cut on the SED. The lower panels show the influence of the initial period models on SED. Black curves (50\,kyr $<\tau_{\rm c}<$ 10\,Myr, $P_0=0.1P$) are the same with the dotted-dashed green curves in Figure~5 of the main text.}
\label{fig:factor_spec}
\end{figure*}

\section{Model for CR-ISM interactions}
We model the CR-ISM interactions following the same method suggested in Refs.~\cite{ZhangR23, Yuan20}. In Ref.~\cite{ZhangR23}, the authors explored two setups of the CR propagation model: one considering diffusion and convection (DC) of CRs; the other considering diffusion and reacceleration (DR) of CRs. The predicted DGE spectra of the two CR propagation models only have slight differences below $\sim 10$\,GeV. Here we adopt the DC model and corresponding parameters, as described in Table 1 of Ref.~\cite{ZhangR23}. In addition to the generated pionic gamma-ray spectrum as already shown in Ref.~\cite{ZhangR23}, we also present the all-sky template (the top-left panel of Supplementary Fig.~\ref{fig:crgam_profile}) and the one after the mask (the top-right panel of Supplementary Fig.~\ref{fig:crgam_profile}). We then obtain the 1D Galactic longitudinal profile in the energy ranges of $10-63\,$TeV and $63-1000\,$TeV, extracted from the same ROI of LHAASO's DGE analysis. The results are compared with LHAASO's measurement in the bottom panels of Supplementary Fig.~\ref{fig:crgam_profile}, based on which the flux ratio presented in Fig.6 is obtained.

\begin{figure*}[htbp]
\centering
\includegraphics[width=\textwidth]{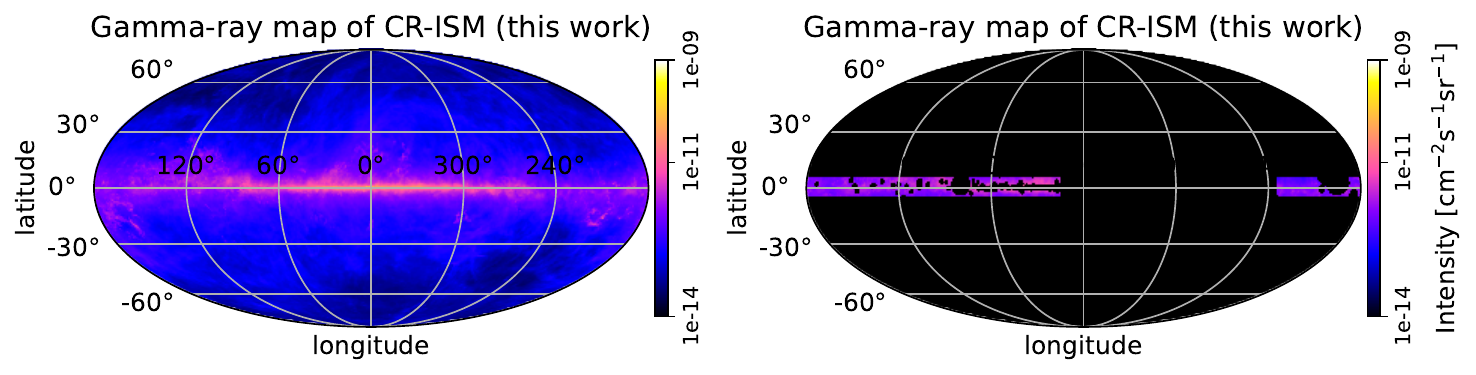}\\
\includegraphics[width=\textwidth]{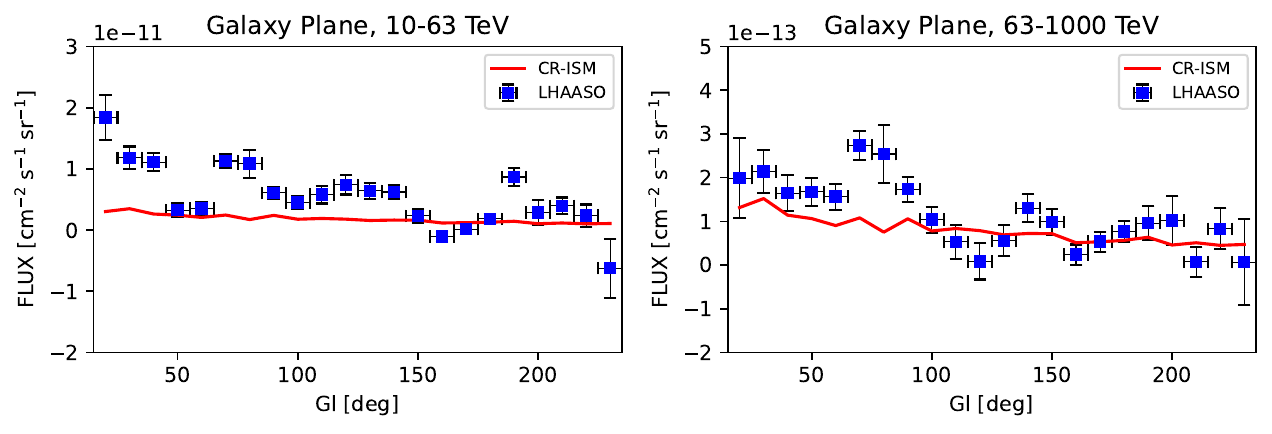}
\caption{All-sky gamma-ray intensity map and Galactic longitudinal gamma-ray profile predicted by the CR-ISM interaction model considered in this work. Top panels: All-sky gamma-ray intensity map in $10-1000$\,TeV predicted by the CR-ISM interaction model considered in this work (top-left) and the one keeping only the ROI of LHAASO's DGE analysis (top-right). Bottom panels: Galactic longitudinal gamma-ray profile from CR-ISM interactions (red curves) compared with the measurements of LHAASO (blue squares, error bars indicating the 1$\sigma$ uncertainties), in the range of $10-63\rm \,TeV$ (bottom-left) and $63-1000\rm \,TeV$ (bottom-right).}
\label{fig:crgam_profile}
\end{figure*}

\begin{figure*}[htbp]
\includegraphics[width=0.5\textwidth]{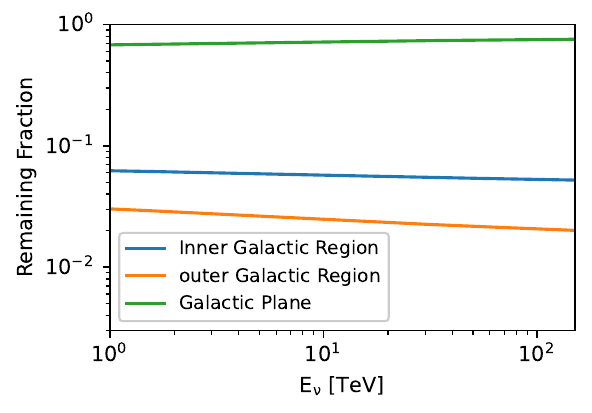}\\
\caption{Remaining fraction of the neutrino flux in the ROI of LHAASO's analysis to the all-sky flux in the KRA$_\gamma^5$ template. The green curve shows that in the entire Galactic plane $|b| \le 5^{\circ}$. The blue and orange curve show the inner Galactic region of $|b| \le 5^{\circ}$, $15^\circ<l<125^\circ$ and outer Galactic region of $|b| \le 5^{\circ}$, $125^\circ<l<235^\circ$, respectively.}
\label{fig:ratio_kragamma}
\end{figure*}

\clearpage

\noindent {\bf Reference}\\
\bibliographystyle{naturemag}

\end{document}